\documentclass[conference]{IEEEtran}
\usepackage{cite}

\ifCLASSINFOpdf
  \usepackage{graphicx}
  \usepackage{epstopdf}
\else
  \usepackage[dvips]{graphicx}
\fi
\DeclareGraphicsExtensions{.eps}
\usepackage[cmex10]{amsmath}
\usepackage{url}

\usepackage{hyperref}

\DeclareMathOperator{\links}{links}

\DeclareMathOperator{\KNassoc}{KNassoc}
\DeclareMathOperator{\diag}{diag}

\begin{document}
%
\title{Revealing Social Networks of Spammers Through Spectral Clustering}



%
\author{\IEEEauthorblockN{Kevin S. Xu\IEEEauthorrefmark{1},
Mark Kliger\IEEEauthorrefmark{2},
Yilun Chen\IEEEauthorrefmark{1}, 
Peter J. Woolf\IEEEauthorrefmark{1}, and
Alfred O. Hero III\IEEEauthorrefmark{1}}
\IEEEauthorblockA{\IEEEauthorrefmark{1}University of Michigan, Ann Arbor,
MI 48109 USA\\
\IEEEauthorrefmark{2}Medasense Biometrics Ltd., PO Box 633, Ofakim, 87516
Israel\\
\IEEEauthorrefmark{1}\{xukevin,yilun,pwoolf,hero\}@umich.edu,
\IEEEauthorrefmark{2}mark@medasense.com}}


\maketitle

\begin{abstract}
To date, most studies on spam have
focused only on the spamming phase of the spam cycle and have ignored the
harvesting phase, which consists of the mass acquisition of email addresses.
It has been observed that spammers conceal their identity to a lesser degree
in the harvesting phase, so it may be possible to gain new
insights into spammers' behavior by studying the behavior of harvesters,
which are individuals or bots that collect email addresses.
In this paper, we reveal social
networks of spammers by identifying communities of harvesters with high
behavioral similarity using spectral
clustering. The data analyzed was collected through Project Honey Pot,
a distributed system
for monitoring harvesting and spamming. Our main findings are (1) that most
spammers either send only
phishing emails or no phishing emails at all, (2) that most communities of
spammers also send only
phishing emails or no phishing emails at all, and
(3) that several groups of spammers within
communities exhibit coherent temporal behavior and have
similar IP addresses. Our findings reveal some previously unknown behavior
of spammers and suggest that there is indeed social structure between spammers
to be discovered.
\end{abstract}


%
\IEEEpeerreviewmaketitle

\section{Introduction}
\begin{figure}[!t]
\centering
\includegraphics[width=3.4in]{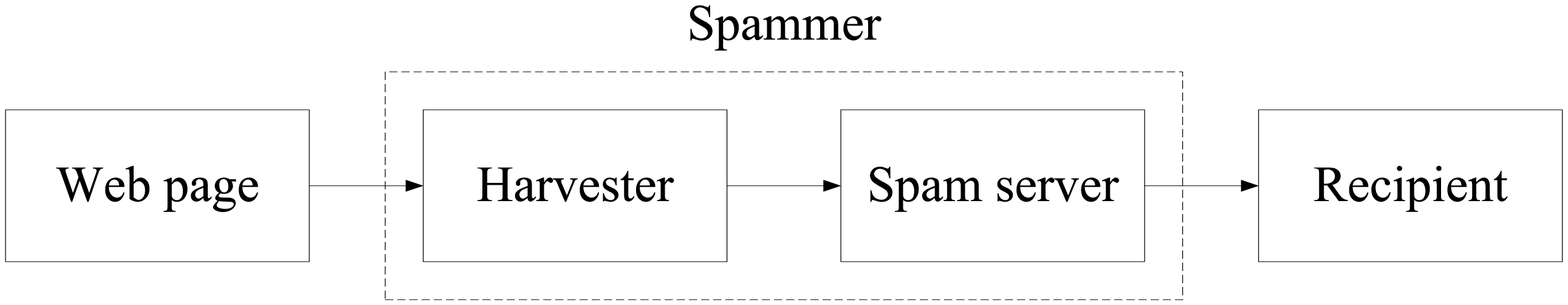}
\caption{The path of spam: from an email address on a web page to a
recipient's inbox.}
\label{fig:SpamPath}
\end{figure}

Previous studies on spam have mostly focused on studying its content or its
source. Likewise, currently used anti-spam methods mostly involve filtering
emails based on their content or by their email server IP address. More
recently, there have been studies on the network-level behavior of spammers
\cite{Ramachandran:SIGCOMM2006,Duan:ICC2007}. However, very little attention
has been devoted to studying how spammers acquire the email addresses that
they send spam to, a process commonly referred to as harvesting. Harvesting is
the first phase of the spam cycle; sending the spam emails to the acquired
addresses is the second phase. Spammers send spam emails using spam servers,
which are typically compromised computers or open proxies, both of which allow
spammers to hide their identities. On the other hand, it has been observed
that spammers do not make the same effort to
conceal their identities during the harvesting phase \cite{Prince:CEAS2005},
indicating that harvesters, which are individuals or bots that collect email
addresses, are closely related to the spammers who are sending the spam
emails. The harvester and spam server are the two intermediaries in the path
of spam, illustrated in Fig. \ref{fig:SpamPath}.

In this paper we try to reveal social networks of spammers by identifying
communities of harvesters using data from
both phases of the spam cycle.
The source of the data analyzed in this paper
is Project Honey Pot \cite{ProjectHoneyPot}, a web-based network for
monitoring harvesting and spamming activity by using trap email addresses.
For every spam email received at a trap email address, the Project Honey Pot
data set
provides us with the IP address of the harvester that acquired the recipient's
email address in addition to the
IP address of the spam server, which is contained in the header of the email.
Spammers make
use of both harvesters and spam servers in order to distribute emails to
recipients, but the IP address of the harvester that acquired the
recipient's email address is typically unknown; it is
only through Project Honey Pot that we are able to uncover it. The Project
Honey Pot data set is described in detail in Section \ref{sec:ProjectHoneyPot}.

Project Honey Pot happens to be an ideal data source for studying phishing
emails. Phishing is an attempt to fraudulently acquire sensitive information
by appearing to represent a trustworthy entity. It is impossible for a trap
email address to, for example, sign up for a PayPal account, so all emails
supposedly received from financial institutions can immediately be classified
as phishing emails. We investigate the prevalence of phishing emails and their
distribution among harvesters.

We look for community structure within the network of harvesters by
partitioning harvesters into groups such that the harvesters in
each group exhibit high behavioral similarity. This is a clustering problem,
and we adopt a method commonly referred to as spectral clustering.
Identifying community structure not only
reveals groups of harvesters that have high behavioral similarity but also
groups
of spammers who may be socially connected, due to the close relation between
harvesters and spammers. We provide an overview
of spectral clustering in Section \ref{sec:Overview}, and we discuss our
choices of behavioral similarity measures in Section \ref{sec:Methodology}.

Our main findings are as follows:
\begin{enumerate}
\item \emph{Most harvesters are either phishers or non-phishers (Section
\ref{sec:ProjectHoneyPot}).} We find that most harvesters either send only
phishing emails or no phishing emails at all (we define what it means for a
harvester to send an email in Section \ref{sec:ProjectHoneyPot}).

\item \emph{Phishers and non-phishers tend to separate into different
communities when clustering based on similarity in spam server usage (Section
\ref{sec:ResultsSpamServers}).} That is, phishers tend to associate with other
phishers, and non-phishers tend to associate with other non-phishers. In
particular, phishers appear in small communities with strong ties, which
suggests that they are sharing resources (spam servers) with other members of
their community.

\item \emph{Several groups of harvesters have coherent temporal behavior and
similar IP addresses (Section \ref{sec:ResultsTemporalSpamming}).}
In particular, we identify a group of ten harvesters that send extremely large
amounts of spam and have the same /24 IP address prefix, which happens to
be owned by a rogue Internet service provider. This indicates that these
harvesters are either the same spammer or a group of spammers
operating from the same physical location.
\end{enumerate}
These findings suggest that spammers do indeed form social networks, and we
are able to identify meaningful communities.

\section{Project Honey Pot}
\label{sec:ProjectHoneyPot}
Project Honey Pot is a distributed system for monitoring harvesting and
spamming activity via a network of decoy web pages with trap email addresses,
known as honey pots. These trap addresses are embedded within the HTML source
of a web page and are invisible to human visitors. Spammers typically acquire
email addresses either by browsing web sites and looking for them or by
running automated harvesting bots that scan
the HTML source of web pages and collect email addresses automatically. Since
the trap email addresses in the honey pots are invisible to human visitors,
Project Honey Pot is trapping only the harvesting bots, and as a
result, this is the only type of harvester that we investigate in this paper.

Each time a harvester visits a honey pot, the centralized Project Honey Pot
server
generates a unique trap email address. The harvester's IP address is recorded
and sent to the Project Honey Pot server. The email address embedded into each
honey pot is
unique, so a particular email address could only have been collected by the
visitor to that particular honey pot. Thus, when an email is received at one
of the trap addresses, we know exactly who acquired the address.
These email addresses are not published anywhere besides the honey pot, so
we can assume that all emails received at these addresses are spam.

\begin{figure}[!t]
\centering
\includegraphics[width=2.5in]{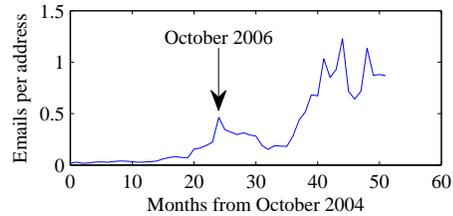}
\caption{Number of emails received by month per address collected.}
\label{fig:EmailsByMonth}
\end{figure}

As of
February 2009, over $35$ million trap email addresses, $39$ million spam
servers, and $59,000$ harvesters have been identified by Project Honey Pot
\cite{ProjectHoneyPot}. Honey pots are located in over $119$ countries.
The total number
of emails received at the trap email addresses monitored by Project Honey Pot
is shown by month in Fig. \ref{fig:EmailsByMonth}, starting from its inception
in October 2004. The number of emails
received have been normalized by the number of addresses collected to
distinguish
between growth of Project Honey Pot and an increase in spam volume.
October 2006 is a month of particular interest. Notice that the
number of emails received in October 2006 increased significantly from
September 2006 then came back down in November 2006. This observation agrees
with media reports of a spam outbreak in October 2006 \cite{ITPro:Spam2006};
thus we will focus our analysis around this time.
We refer readers to \cite{Prince:CEAS2005, ProjectHoneyPot} for additional
details on Project Honey Pot.

In order to discover social networks of spammers, we need to
associate emails to the spammers who sent them. Since we do not
know the identity of the spammer who sent a particular email, we can associate
the email either to the spam server
that was used to send it or to the harvester that acquired the recipient's
email address. A previous study
using the Project Honey Pot data set has suggested that the harvester is more
likely to be associated with the spammer than the spam server
\cite{Prince:CEAS2005}, so \emph{we associate each email with the harvester
that acquired the recipient's email address}. In particular, this is
different from studies \cite{Ramachandran:SIGCOMM2006, Duan:ICC2007},
which did not involve harvesters and implicitly associated emails with
the spam servers that were used to send them.
Note that we are not assuming that the harvesters are
the spammers themselves. A harvester may collect email addresses for multiple
spammers, or a spammer may use multiple harvesters to collect email
addresses. Thus, when we say that a particular harvester sends an email, we
mean that a \emph{spammer who obtained an address collected by this harvester
sends an email}.
To summarize, we are associating emails with harvesters and trying
to discover communities of harvesters, which are closely related
to communities of actual spammers.

As mentioned previously, Project Honey Pot is an ideal data source for
studying phishing emails because the trap email addresses cannot sign up
for accounts at financial institutions and other sources that phishing emails
fraudulently represent. Note that this is not possible with legitimate email
addresses, which may receive legitimate emails from these sources. Since we
know that any email mentioning such a source is a phishing email, we can
classify each email as phishing or non-phishing based on its content. We
classify an email as phishing if its subject contains a commonly used phishing
word. The list of such words was built using common phishing words such as
``password'' and ``account'' and includes
those found in a study on phishing \cite{Chandrasekaran:2006} and names
of large
financial institutions that do business on-line such as PayPal and Chase.

In general we find that a small percentage of the spam received through
Project Honey Pot consists of phishing emails. As of
February 2009, $3.5\%$ of
the spam received was phishing spam. We define a phishing
level for each harvester as the ratio of the number of phishing emails it
sent to the total number of emails it sent. An interesting finding is that
\emph{most harvesters either send only phishing emails or no phishing emails
at all}.
In particular, $14\%$ of harvesters have
a phishing level of $0.9$ or higher while $77\%$ have a phishing level of $0.1$
or lower, with only $9\%$ of harvesters in between.
Thus we can label all
harvesters as phishers or non-phishers based on their phishing level. We
label a harvester as a phisher if its phishing level exceeds $0.5$. As of
February 2009, about $18\%$ of harvesters were labeled as phishers.
We note that phishers
send less emails on a per-harvester basis than non-phishers, as only $3.5\%$ of
emails received were phishing emails as mentioned earlier. The labeling of
harvesters as phishers or non-phishers will be used later when interpreting
the clustering results.

\section{Overview of Spectral Clustering}
\label{sec:Overview}
In this paper, we employ spectral clustering
to identify groups of harvesters with high behavioral similarity. We choose
spectral clustering over other
clustering techniques because of its close relation to the graph partitioning
problem of minimizing the normalized cut between partitions, which is a
natural choice of objective function for community detection as discussed in
\cite{Leskovec:WWW2008} where it is referred to as conductance.

\subsection{The graph partitioning problem}
We represent the network of harvesters by a weighted undirected graph $G =
(V,E,W)$ where $V$ is the set of vertices, representing harvesters; $E$ is the
set of edges between vertices; and $W = \left[w_{ij}\right]_{i,j=1}^M$ is the
matrix of edge weights with $w_{ij}$ indicating the similarity
between harvesters $i$ and $j$. The choice of similarities is discussed in
Section \ref{sec:Methodology}.
$W$ is the adjacency matrix of the graph and
is also referred to in the literature as the similarity matrix or affinity
matrix. $M=|V|$ is the total number of harvesters. The total weights of edges
between two sets of vertices $A,B \subset V$ is defined by
\begin{equation}
\links(A,B) = \sum_{i \in A} \sum_{j \in B} {{w_{ij}}},
\end{equation}
and the degree of a set $A$ is defined by
\begin{equation}
\deg(A) = \links(A,V).
\end{equation}

Our objective is to find highly similar groups of vertices in the graph,
which
represent harvesters that behave in a similar manner. This is a graph
partitioning problem, and our objective
translates into minimizing similarity between groups, maximizing similarity
within groups, or preferably both. Let the groups be denoted by $V_1, V_2,
\ldots, V_K$ where $K$ denotes the number of groups to partition the graph
into. We represent the graph partition by an $M$-by-$K$ partition matrix $X$.
Let $X = [\mathbf{x_1}, \mathbf{x_2}, \ldots, \mathbf{x_K}]$ where
$x_{ij} = 1$ if harvester $i$ is in cluster $j$ and $x_{ij} = 0$ otherwise.
We adopt the normalized cut disassociation measure proposed in
\cite{Shi:PAMI2000}.
One favorable property of this measure is that minimizing the normalized
cut between groups simultaneously maximizes the
normalized association within groups. Thus we
attempt to minimize the normalized cut by maximizing the normalized
association within groups, which is defined by
\begin{equation}
\KNassoc(X) = \frac{1}{K} \sum_{i=1}^K {\frac{\links(V_i, V_i)}{\deg(V_i)}}.
\end{equation}

\subsection{Finding a near global-optimal solution}
Unfortunately, maximizing $\KNassoc$ is NP-complete even for $K=2$ as noted in
\cite{Shi:PAMI2000} so we turn to an approximate method. Define the degree
matrix $D = \diag(W1_M)$ where $\diag(\cdot)$ creates a diagonal matrix from
its vector argument, and $1_M$ is a vector of $M$ ones. Rewrite $\links$ and
$\deg$ as
\begin{align}
\links(V_i, V_i) & = \mathbf{x_i}^T W \mathbf{x_i} \\
\deg(V_i) & = \mathbf{x_i}^T D \mathbf{x_i}.
\end{align}
We can formulate the $\KNassoc$ maximization problem as follows:
\begin{align}
\textrm{maximize} \quad & \KNassoc(X) = \frac{1}{K} \sum_{i=1}^K
{\frac{\mathbf{x_i}^T W \mathbf{x_i}}{\mathbf{x_i}^T D \mathbf{x_i}}} \\
\textrm{subject to} \quad & X \in \{0,1\}^{M \times K} \\
& X1_K = 1_M.
\end{align}
As mentioned earlier, finding the optimal partition matrix $X$ is an
NP-complete problem. A near
global-optimal solution can be found by first relaxing a transformed version
of $X$ into the continuous domain and finding the optimal continuous partition
matrix by solving a generalized eigenvalue problem. This is followed by
solving a discretization problem where the closest discrete partition matrix
to the optimal continuous partition matrix is sought. We refer interested
readers to \cite{Yu:ICCV2003} for details on this method, commonly referred
to as spectral clustering.

\subsection{Choosing the number of clusters}
As with most clustering algorithms, the proper choice of $K$, the number of
clusters, is unknown in spectral clustering. A useful heuristic particularly
well-suited for
choosing $K$ in spectral clustering problems is the eigengap heuristic. The
goal is to choose $K$ such that the highest eigenvalues $\lambda_1, \ldots,
\lambda_K$ of
the adjacency matrix $W$ are very close to $1$ but $\lambda_{K+1}$ is
relatively far away from $1$. This
procedure was justified in \cite{vonLuxburg:2007} and is used to choose $K$ in
this paper.

\section{Methodology}
\label{sec:Methodology}
A social network is a social structure composed of nodes, also known as
actors, and ties, which indicate the relationships between nodes. We cannot
observe direct relationships between harvesters (the actors), so we use
indirect relationships as the ties. We explore two types of
ties in this paper. Each type of tie corresponds to a similarity measure for
choosing the edge weights $w_{ij}$, which indicate the behavioral similarity
between harvesters.

Note that the network may evolve over time so we need to choose a time frame
for analysis that is short enough so that we should be able to see this
evolution if it is present yet long enough so that we have a large enough
sample for the clustering results to be meaningful. There is no clear-cut
method for choosing the time frame. As a starting point, we split
the data set by month and
analyze each month independently.

\subsection{Similarity measures}
In this paper, we study two measures of behavioral similarity: similarity in
spam server usage and temporal similarity.
For both of these similarity measures, we create a coincidence
matrix $H$ as an intermediate step to the creation of the adjacency matrix
$W$, which is discussed in Section \ref{sec:CreatingAdjMatrix}. The choice of
similarity measure is crucial because it determines the topology of the graph.
Each similarity measure provides a different view of the social network, so a
poor choice of similarity measure may lead to detecting no community structure
if harvesters are too similar or too dissimilar.

\subsubsection{Similarity in spam server usage}
We note that harvesters typically send emails through multiple spam servers so
common usage of spam servers is one way to link harvesters. Consider a mixed
network of harvesters and spam servers described by the $M \times N$
coincidence
matrix $H = \left[h_{ij}\right]_{i,j=1}^{M,N}$, where $M$ is the number of
harvesters and $N$ is the number of spam servers. We choose $h_{ij} =
p_{ij}/\left(d_j e_i\right) \in [0,1]$ where $p_{ij}$ denotes the number of
emails sent by harvester $i$ using spam server $j$, $d_j$ denotes the total
number of emails sent (by all harvesters) through spam server $j$, and $e_i$
denotes the total
number of email addresses harvester $i$ has acquired. $d_j$ is a normalization
term that is included to account for the variation in the total number of
emails sent through each spam server. For example, a harvester that sent four
emails through a spam server which only sent four emails total should indicate
a much stronger connection to that spam server than one that sent four emails
through a spam server which sent one thousand emails total. $e_i$ is also a
normalization term to account for the variation in the number of email
addresses each harvester has acquired, based on the assumption that harvesters
send an equal amount of spam to each address they have acquired.
We can interpret $h_{ij}$ as \emph{harvester $i$'s
percentage
of usage of spam server $j$ per address it has acquired}. The similarity
between two harvesters $i_1$ and $i_2$ is the inner product between rows
$i_1$ and $i_2$ of $H$.

\subsubsection{Temporal similarity}
Harvesters that exhibit high similarity in their temporal behavior may also
indicate a social connection, so another possibility for linking harvesters is
by their temporal spamming behavior. We look at the timestamps of all emails
sent by a particular harvester and bin them into $1$-hour intervals, resulting
in
a vector indicating how many emails a harvester sent in each interval. Doing
this for all of the harvesters, we get another coincidence matrix $H$ but with
the columns representing time (in $1$-hour intervals) rather than spam servers.
The entries of $H$ are $h_{ij} = s_{ij}/e_i$ where $s_{ij}$ denotes the number
of emails sent by harvester $i$ in the $j$th time interval, and $e_i$
is defined as
before. Again we normalize by the number of email addresses acquired but no
other normalizations are necessary because the columns represent time, which
does not vary for different harvesters.

\subsection{Creating the adjacency matrix}
\label{sec:CreatingAdjMatrix}
From the coincidence matrix $H$ we can obtain an unnormalized matrix of
pairwise similarities $S = HH^T$. We normalize $S$ to form a normalized matrix
of pairwise similarities $S' = D_S^{-1/2}SD_S^{-1/2}$, where $D_S$ is a
diagonal matrix consisting
of the diagonal elements of S. We can interpret this final
normalization as a scaling of the edge weights between harvesters such that
each harvester's self-edge has unit weight. This ensures that each harvester
is equally
important because we have no prior information on the importance of a
particular harvester in the network.

We create an adjacency matrix $W$ describing the graph by connecting the
harvesters together according to their similarities in $S'$. There are several
methods of connecting the graph, including $k$-nearest neighbors and
the fully-connected graph. We opt for the $k$-nearest
neighbor method, which translates into connecting each node to its neighbors
with the $k$ highest similarities. This is the recommended choice in
\cite{vonLuxburg:2007} and is less vulnerable to improper choices of the
connection parameters (in this case, the value of $k$).
It also results in a sparse adjacency matrix,
which speeds up computations and makes the graph easier to
visualize. Unfortunately, there are not many guidelines on how to choose $k$.
A heuristic suggested in \cite{vonLuxburg:2007}, motivated by asymptotic
results, is to choose $k = \log M$. We use this choice of $k$ as a starting
point and increase $k$ as necessary to avoid artificially disconnecting the
graph.

\section{Results}
\label{sec:Results}
We present visualizations for our clustering results from October 2006, which
is a month of particular interest as noted in Section
\ref{sec:ProjectHoneyPot}. The visualizations were created
using the force-directed layout in Cytoscape \cite{Cytoscape}. Key statistics
of the clustering results over a period of one year starting in July 2006
are presented in $3$-month intervals in tables.

\subsection{Similarity in spam server usage}
\label{sec:ResultsSpamServers}
\begin{figure}[!t]
\centering
\includegraphics[width=2.9in]{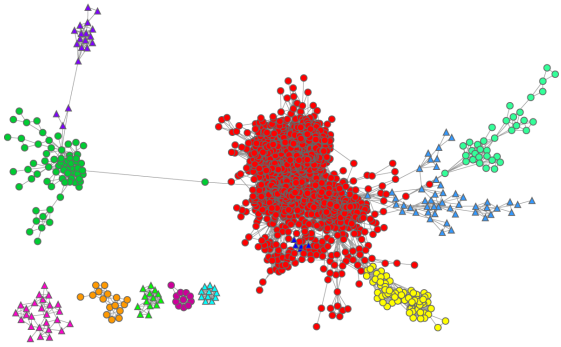}
\caption{Social network of harvesters formed by similarity in spam server
usage in October 2006 (best viewed in color). The color and shape of a
harvester indicate the cluster it belongs to.}
\label{fig:200610SSClusters}
\end{figure}

The graph created using similarity in spam server usage usually consists of
a large connected component and many small connected components.
The small components are easily
recognized as clusters, while the large component is divided into multiple
clusters. In Fig. \ref{fig:200610SSClusters} we show the social
network of harvesters, connected using similarity in spam server usage, from
October 2006. The shape and color of a harvester indicates the cluster
it belongs to. The eigengap heuristic suggests that the large connected
component should be divided into $64$ clusters, but to make the figure easier
to interpret, we present a clustering result that divides the large component
into $7$ clusters. We also remove connected components of less than ten
harvesters. These modifications were made for visualization
purposes only. In our analysis, and in particular when calculating the
validation indices we present later, we use the number of clusters suggested
by the eigengap heuristic and include all small connected components.

Notice that the
majority of harvesters belong in a large cluster with weak ties, which is
a subset of the large component. Meanwhile there exist several smaller
clusters with strong ties, some of which are connected to the large cluster.
Each cluster represents a community of harvesters that happen to use the same
resources (spam servers), indicating that there is a strong likelihood that
these harvesters are working together.

\begin{figure}[!t]
\centering
\includegraphics[width=2.9in]{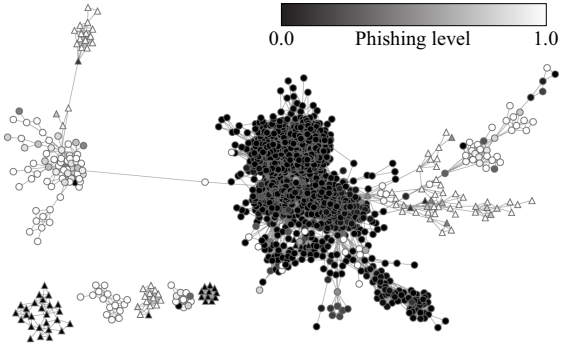}
\caption{Alternate view of social network pictured in Fig.
\ref{fig:200610SSClusters}, where the color of a harvester corresponds to
its phishing level.}
\label{fig:200610SSPhishing}
\end{figure}

As with any clustering problem, the results need to be validated. If common
usage of spam servers indeed indicates social connections between harvesters,
perhaps we can find some other property that is consistent within clusters.
Recall
from Section \ref{sec:ProjectHoneyPot} that harvesters can be classified as
either phishers or non-phishers. In Fig. \ref{fig:200610SSPhishing} we show
the same social network colored by phishing level, as defined in Section
\ref{sec:ProjectHoneyPot}, rather than cluster. Note that each of the clusters
consists almost entirely of phishers or almost entirely of non-phishers.
In particular, phishers appear to concentrate in small clusters with strong
ties.
This observation is further enhanced when clustering using $64$ clusters as
suggested by the eigengap heuristic. Thus,
phishing level appears to be consistent within clusters.
We consider a cluster as a phishing cluster if it contains
more phishers than non-phishers and as a non-phishing cluster otherwise.

Using phisher or non-phisher as a label for each harvester, we compute the
Rand index and adjusted Rand index \cite{Hubert:1985}, both commonly used
indices used for clustering validation. The Rand index is a measure of
agreement between clustering results
and a set of class labels and is given by
\begin{equation}
\text{Rand index} = \frac{a+d}{a+b+c+d}
\end{equation}
where $a$ is the number of pairs of nodes with the same label and in the same
cluster, $b$ is the number of pairs with the same label but in different
clusters, $c$ is the number of pairs with different labels but in the same
cluster, and
$d$ is the number of pairs with different labels and in different
clusters. A Rand index of $0$ indicates complete
disagreement between clusters and labels, and a Rand index of $1$ indicates
perfect agreement. The adjusted Rand index is
corrected for chance so that the range is $[-1,1]$ with an expected index of
$0$ for a random clustering result.

In this clustering problem, the Rand index indicates how
well phishers and non-phishers divide into phishing and non-phishing clusters,
respectively. The adjusted
Rand index indicates how well phishers and non-phishers divide compared to
the expected division that a random clustering algorithm would produce.
Both indices are shown in Table \ref{tab:ValidationIndices} for five months.
Note that the clustering results have excellent
agreement with the labels, and the agreement is much higher than expected by
chance. The division between phishers and non-phishers is not perfect, as
there are some phishers
belonging in non-phishing clusters and vice-versa, but the high adjusted Rand
index indicates that this split is highly unlikely to be due to chance alone.
Hence we have found empirical evidence that
phishers tend to form small communities with strong ties, suggesting that
they share resources (spam servers)
between members of their community.

\begin{table}[!t]
\renewcommand{\arraystretch}{1.2}
\caption{Validation indices for clustering results}
\label{tab:ValidationIndices}
\centering
\begin{tabular}{|c||c|c||c|c|c|}
\hline
Year & \multicolumn{2}{|c||}{2006} & \multicolumn{3}{|c|}{2007}\\
\hline
Month & July & October & January & April & July\\
\hline
Rand index & 0.923 & 0.954 & 0.942 & 0.964 & 0.901\\
\hline
Adj. Rand index & 0.821 & 0.871 & 0.810 & 0.809 & 0.649\\
\hline
\end{tabular}
\end{table}

\newcommand{\avg}{\mathrm{avg}}

\subsection{Temporal similarity}
\label{sec:ResultsTemporalSpamming}

\begin{figure}[!t]
\centering
\includegraphics[width=1.9in]{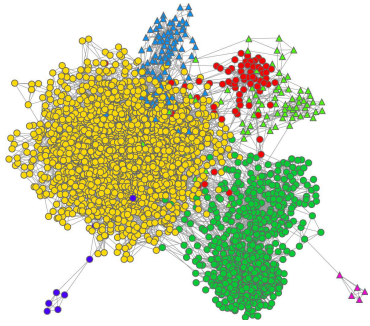}
\caption{Social network of harvesters formed by temporal
similarity in October 2006 (best viewed in color). The color and shape of a
harvester indicate the cluster it belongs to.}
\label{fig:200610TSClusters}
\end{figure}

Unlike the graph created by similarity in spam server usage, the graph created
by temporal similarity is usually connected.
In Fig. \ref{fig:200610TSClusters} we show the social
network of harvesters, connected using temporal similarity, from
October 2006, where again the shape and color of a harvester indicates the
cluster it belongs to. Any similarity in color with Fig.
\ref{fig:200610SSClusters} is coincidental; Fig.
\ref{fig:200610TSClusters} represents a completely different view of the
social network and provides different insights.

Unfortunately we do not have validation for this clustering result on a global
scale like we did with phishing level for similarity in spam server usage.
However by looking at temporal spamming plots of the small clusters,
we find some local validation. Namely, we see groups of harvesters in
the same cluster with extremely coherent temporal spamming
behavior. We notice that in many of these groups,
the harvesters also have similar IP addresses.
In particular, we notice a
group of ten harvesters that have extremely coherent temporal spamming patterns
and have the same /24 IP address prefix, namely 208.66.195/24,
indicating that they are also in the same physical location.
In Fig. \ref{fig:200610TSClusters} they can be found in the light green cluster
of triangular nodes at the top right of the network.

Upon further investigation, we find that their IP addresses
are in the 208.66.192/22 prefix owned by McColo Corp., a known rogue Internet
service provider that acted as a gateway to spammers and was finally removed
from the Internet in November 2008 \cite{ArborNet:McColo2008}. This serves
as further confirmation that these harvesters are likely to be socially
connected. They
first appeared at the end of May 2006 and have been
among the heaviest harvesters, in terms of the number of emails sent, in every
month since then. The average correlation coefficients $\rho_{\avg}$ between
two harvesters in this group are listed in Table
\ref{tab:AvgCorrelationCoefficients} for five months. Notice that their
average correlation coefficients are extremely high and strongly suggest that
they are working together in a coordinated matter. Also note that
their behavior is still highly correlated more than a year after they first
appeared. Furthermore, we discover that they have high temporal correlation in
the harvesting phase; that is, they collect email addresses in a very similar
manner as well. We would certainly expect them to belong to
the same cluster, which agrees with the clustering results. Hence we believe
that this group is either the same spammer or a community of spammers
operating from the same physical location.

\begin{table}[!t]
\renewcommand{\arraystretch}{1.2}
\caption{Average temporal correlation coefficients of 208.66.195/24 group of
ten harvesters}
\label{tab:AvgCorrelationCoefficients}
\centering
\begin{tabular}{|c||c|c||c|c|c|}
\hline
Year & \multicolumn{2}{|c||}{2006} & \multicolumn{3}{|c|}{2007}\\
\hline
Month & July & October & January & April & July\\
\hline
$\rho_{\avg}$ & 0.980 & 0.988 & 0.950 & 0.949 & 0.935\\
\hline
\end{tabular}
\end{table}

\section{Conclusions}
In this paper, we revealed social networks of spammers by discovering
communities of harvesters from the data collected
through Project Honey Pot. Specifically, we clustered harvesters using two
similarity measures reflecting their
behavioral correlations. In addition, we studied the distribution of phishing
emails among harvesters and among clusters. We found that harvesters typically
send either only phishing emails or no phishing emails at all. Moreover,
we discovered
that communities of harvesters divide into communities of mostly phishers and
mostly non-phishers when clustering according to similarity in spam server
usage. In particular, we observed that phishers tend to form small
communities with strong ties. We also
discovered several groups of harvesters with extremely coherent temporal
behavior and very similar IP addresses, indicating that these groups are
close geographically in addition to socially.

Note that the two similarity measures we studied provided us with different
views of the social networks of harvesters,
and we gained useful insights from both of them. All of
our findings are empirical; however, we believe that they reveal some
previously unknown behavior of spammers, namely that
spammers do indeed form social networks. Since harvesters are closely
related to spammers, the discovered communities of harvesters are closely
related to communities of spammers. If we further hypothesize
that harvesters are the spammers themselves, then the discovered communities
of harvesters correspond exactly to communities of spammers. Identifying
communities of spammers allows us to fight spam from a new
perspective---by using spammers' social structure.


\section*{Acknowledgment}
We thank Matthew Prince, Eric Langheinrich, and Lee Holloway of Unspam
Technologies Inc. for providing us with the Project Honey Pot data set. We are
also grateful to Nitin Nayar, John Bell, and Dr. Olaf Maennel for their
assistance with the data retrieval. This research was partially supported by
the Office of Naval Research grant N00014-08-1-1065 and the National Science
Foundation grant CCR-0325571. Kevin Xu was supported in part by an award from
the Natural Sciences and Engineering Research Council of Canada.



\bibliographystyle{IEEEtran}
\bibliography{IEEEabrv,References}
%
%
%

\end{document}